\documentclass[aps,pre,reprint,groupedaddress]{revtex4-1}
 
\usepackage{color,graphicx}
\usepackage{float}
\usepackage{color,amsmath}

\begin{document}

\title{Box-scaling as a proxy of finite-size correlations}

\author{Daniel A. Martin$^{1,6}$}  
\author{Tiago L. Ribeiro$^{2}$}  
\author{Sergio A. Cannas$^{3,6}$}
\author{Tomas S. Grigera$^{4,5,6}$}
\author{Dietmar Plenz$^{2}$}
\author{Dante R. Chialvo$^{1,6}$}
%\thanks{}

\affiliation{$^1$Center for Complex Systems $\&$ Brain Sciences (CEMSC$^3$), Universidad Nacional de San Mart\'in, 
San Mart\'in, (1650) Buenos Aires, Argentina}

\affiliation{$^2$Section on Critical Brain Dynamics, National Institute of Mental Health, National Institutes of Health, Bethesda,
MD, 20892, USA}
\affiliation{$^3$Facultad de Matem\'atica Astronom\'ia F\'isica y Computaci\'on, Universidad Nacional de C\'ordoba, Instituto de F\'isica Enrique Gaviola (IFEG-CONICET), Ciudad Universitaria. (5000) C\'ordoba, Argentina.}

\affiliation{$^4$Instituto de F\'isica de L\'iquidos y Sistemas Biol\'ogicos (IFLySiB), CCT CONICET La Plata and Universidad Nacional de La Plata, Calle 59 no 789, B1900BTE La Plata, Argentina}

\affiliation{$^5$Departamento de F\'isica, Facultad de Ciencias Exactas,  Universidad Nacional de La Plata, Argentina}

\affiliation{$^6$Consejo Nacional de Investigaciones Cient\'ificas y T\'ecnicas (CONICET), Godoy Cruz 2290, Buenos Aires, Argentina}
\date{\today}

\begin{abstract}
The  scaling of correlations as a function of system size provides important hints to understand critical phenomena on a variety of systems. Its study in biological systems offers two challenges: usually they are not of infinite size, and in the majority of cases sizes can not be varied at will.  Here we discuss how finite-size scaling can be approximated in an experimental system of fixed and relatively small size by computing correlations inside of a reduced field of view of various sizes (i.e., ``box-scaling''). Numerical simulations of a neuronal network are used to verify such approximation, as well as the ferromagnetic 2D Ising model. The numerical results support the validity of the heuristic approach, which should be useful to characterize relevant aspects of critical phenomena in biological systems.
\end{abstract}
%\keywords{}

\maketitle
 
Complex biological phenomena at all levels including macroevolution, neuroscience at different scales, and molecular biology are of increasing interest. In some systems, the origin of such complexity has been  traced to critical phenomena via models and theory \cite{mora,bak,chialvo,haimovici,reviewBiophys,beggs,tang,flocks,miguel,haimovici,fraiman,PP3}. Nonetheless, the connection between complexity and criticality still needs to be established carefully in each case. Among others, a very distinctive indicator of the presence of critical phenomena is the observation of an increase in the correlation length as a function of the size of the system under study \cite{diffusive, criticaldynamics,DOMB,cavagna2014}. Such observation 
exposes one of the hallmarks of criticality: a complex dynamic which lacks a characteristic scale. Less evident but equally relevant is the fact that  at criticality the only scales are the ones ``imposed'' from outside, i.e.,  the finite size of the system and the limited time of system observation.

In most physical systems, criticality can be studied through the variation of system properties as some external parameter (say temperature) is changed. However this kind of tuning is usually off-limits in biological systems. Alternatively,  one can in principle establish the lack of an intrinsic scale by demonstrating the size-scaling directly, i.e.\ by studying the correlation function of systems of increasing size.  While this can be done with relative ease in numerical studies, it is much harder to achieve in experiments \cite{Cardy}.  This is especially true in  biological systems, which  in most cases can neither be cut in small pieces, nor can they easily be enlarged.  The brain is a prototypical biological system for which critical dynamics has been suggested to hold the key to its core functions. \cite{beggs, chialvo,haimovici,fraiman,PP3}. 
The correlations reported for ongoing and evoked brain activity have been found to depend in a unique way on the size of the observation window \cite{fraiman,haimovici,tiago2020}. However, there has been no systematic analysis whether and how system-size scaling can be approximated by varying the size of an observation window (without changing a system's size). We will refer to this latter approach as ``box-scaling'', since it resembles the fractal box-counting algorithm \cite{box}.

The letter is organized as follows. First  the connected correlation function (CCF) is defined. Then the CCF is studied for a neuronal network model under  two scenarios:  in the first one we proceed in the standard manner, increasing the system size and determining its correlation behaviour. In the second setting, the CCF is examined using a fixed  system size (relatively large) while varying the size of the field of view (changing the observable window size).  After that similar calculations are described in the ferromagnetic 2D Ising model. The letter concludes with a summary of the main results.
 
\emph{Correlation analysis:}  The connected correlation function measures how a local quantity loses spatial correlation as distance is increased \cite{goldenfeld}.  Here we compute it as \cite{cavagna2014,flocks}
\begin{equation}
C(r)= {1 \over c_0} { \sum_{i,j}  u_i u_j \delta(r-r_{ij})  \over \sum_{i,j}  \delta(r-r_{ij})  } 
\label{eq:9}
\end{equation}
where $\delta(r-r_{ij})$  is a smoothed Dirac $\delta$ function selecting pairs of neurons (or region of interest) at mutual distance $r$; $r_{ij}$ is the Euclidean distance from the site $i$ to site $j$;  $u_i$ is the value of the signal $s$ of site $i$ at time $t$, after subtracting the overall mean of signals $s$ at that time $t$, $u_i(t)=s_i(t)-\overline{s}(t)$; ${1 \over c_0}$ is a normalization factor to ensure that $C(r=0)=1$.

The definition Eq.~\eqref{eq:9} of the CCF differs from the statistical mechanic definition of CCF (or its normalized counterpart, the Pearson correlation coefficient) in an important aspect: the $u_i$ are the fluctuations with respect to the \emph{instantaneous space average}, $\overline{s}(t) = (1/N)\sum_i^N s_i(t)$, as opposed to the ensemble, or phase, average $\langle s_i\rangle$, defined through an appropriate probability distribution.  We also note that the effect of the denominator is to compensate for the density fluctuations, i.e.\ the lattice spatial structure, thus disentangling the intrinsic fluctuations of $s_i$ from those due to the discrete distribution of particles in space.  It is particularly convenient when the effects of open boundaries are expected to be relevant  \cite{flocks}. 

%%%%%%%%%%%%%%%%%%FIGURE 1 %%%%%%%%%%%%%%%%%%%
\begin{figure*}[ht!]
 \includegraphics [width = .62 \linewidth] {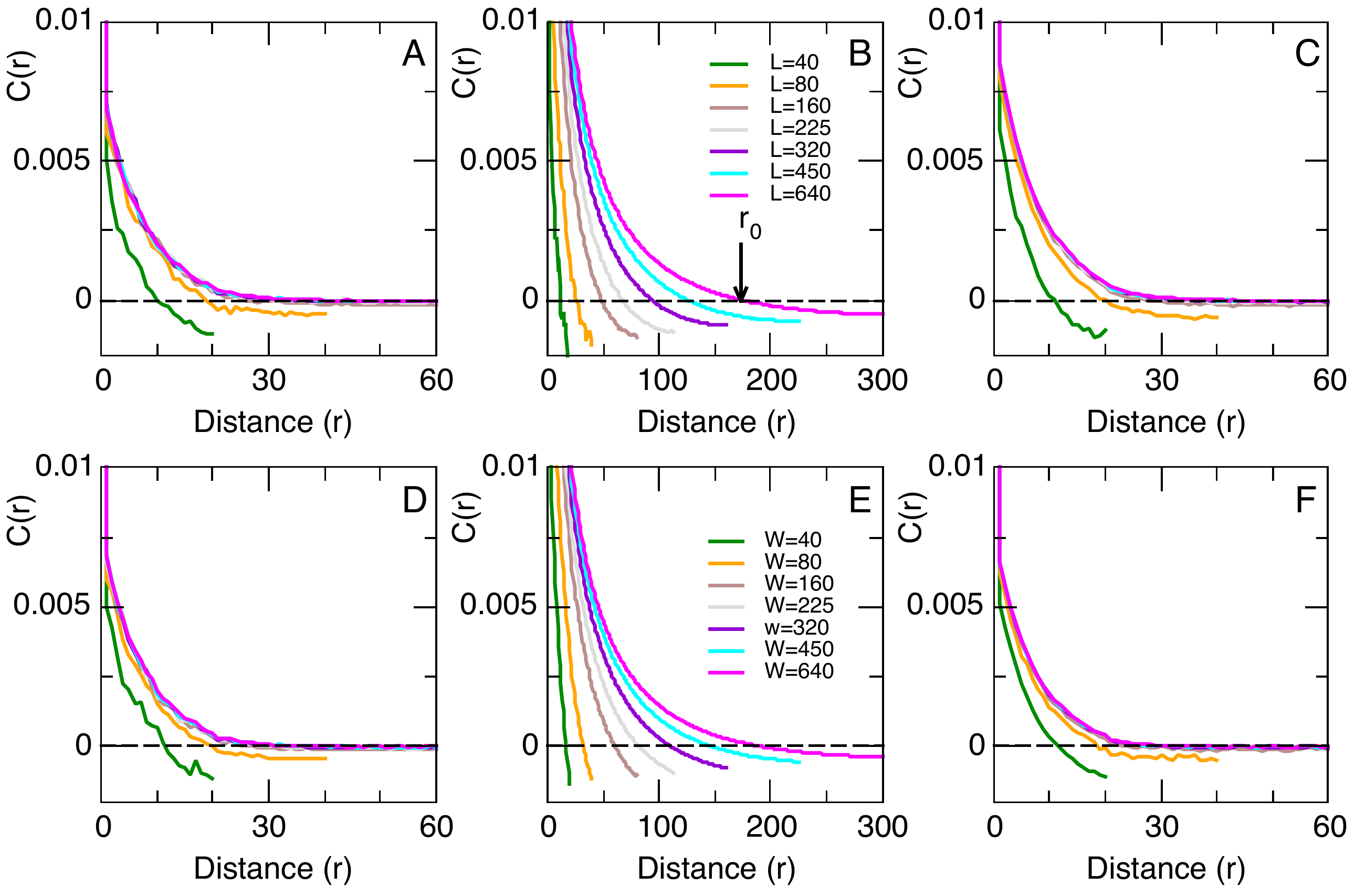}
 \caption{\textbf{Connected correlation function of the neuronal network model}. Curves in panels  A, B and C  for different system sizes $L$ and those in D, E and F for different window sizes $W$ computed on a system of size $L=1000$.  Results are for three control parameter values corresponding to sub-critical ($\sigma$=0.64, panels A and D), critical ($\sigma$=1.024, panels B and E) and super-critical ($\sigma= 1.6$, panels C and F) regimes of the model. Arrow in panel B illustrates the value of $r_0$ for $L=640$.}
\label{F1}
\end{figure*}
%%%%%%%%%%%%%%%%%%%%%%%%%%%%%%%%%%  
%%%%%%%%%%%%%%%%%Figure 2 %%%%%%%%%%%
\begin{figure*}[ht!]
\includegraphics [width = .62 \linewidth] {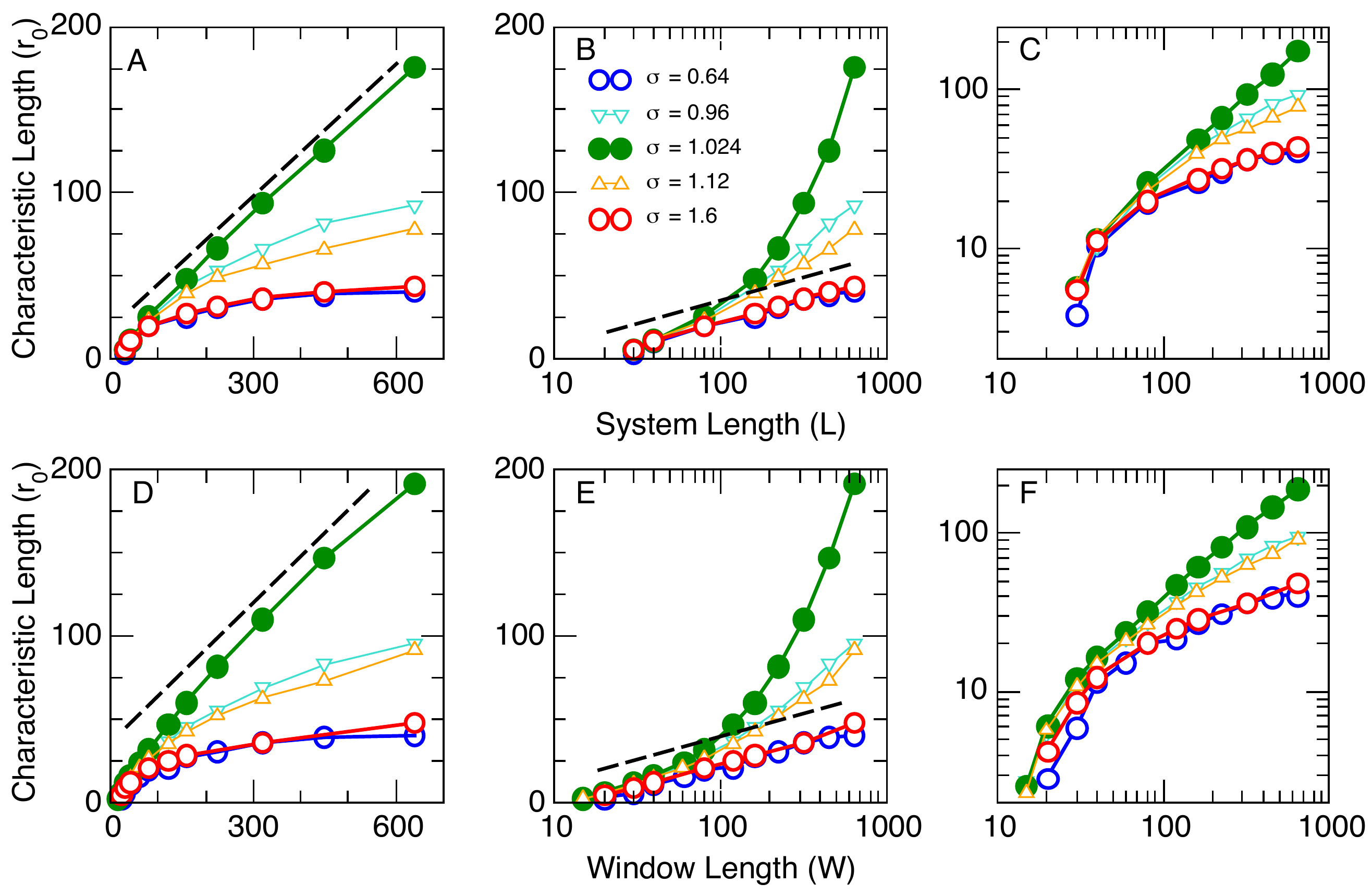}
 \caption{\textbf{Characteristic Length $r_0$ of the neuronal network model:} The zero crossings of the CCF shown in Fig.\ref{F1} are plotted in linear-linear (left), log-linear(middle) and log-log (right) axis. Top three panels correspond to different  system size $L$ and bottom three panels to different box length $W$.  Different symbols correspond to the values of the control parameter $\sigma$ denoted in the legends. Dashed lines are visual aids to emphasize the predicted logarithmic behaviour for both sub-critical and super-critical regimes (open circles) and the linear dependence expected for the critical regime (filled circles). Open triangles are used to denote results obtained for intermediate values of $\sigma$. }
\label{F2}
\end{figure*}
%%%%%%%%%%%%%%%%%%%%%%%%%%%%%%%%%%%

The correlation length $\xi$ measures the scale at which two points start to become uncorrelated.  In a critical system the correlation length is infinite, meaning that the decay of the correlation lacks a characteristic scale.  In this case, the position of the zero of $C(r)$ in Eq.~1 is a useful length scale, because then it increases proportionally to the system size $L$ (the CCF defined with instantaneous space averages substracted \emph{must} have a zero \cite{flocks}).  This  functional dependence, attesting scale invariance, suggests the presence of critical dynamics.  However, it can not be used  when the system size is fixed, as in the case of brain networks. 
Instead, we can define a  \emph{characteristic scale} $r_0$ defined by the first zero of $C(r)$, but with the correlation function computed for partial regions of the entire system. For that the system is subdivided in boxes of side $W$, (with $W < L$) and $r_0$ determined by $C_W(r_0)=0$, where $C_W(r)$ is the CCF Eq.~1 restricted to a box of size $W$. Thus, the implementation follows the same logic and limitations than the box-counting algorithm commonly used to compute the fractal dimension of a data set, image or object \cite{box}.
%%%%%%%%%%%%%%%%%%%%%%%%%FIGURE 3 %%%%%%%%%%
\begin{figure*}[ht!]
\includegraphics [width = .62 \linewidth] {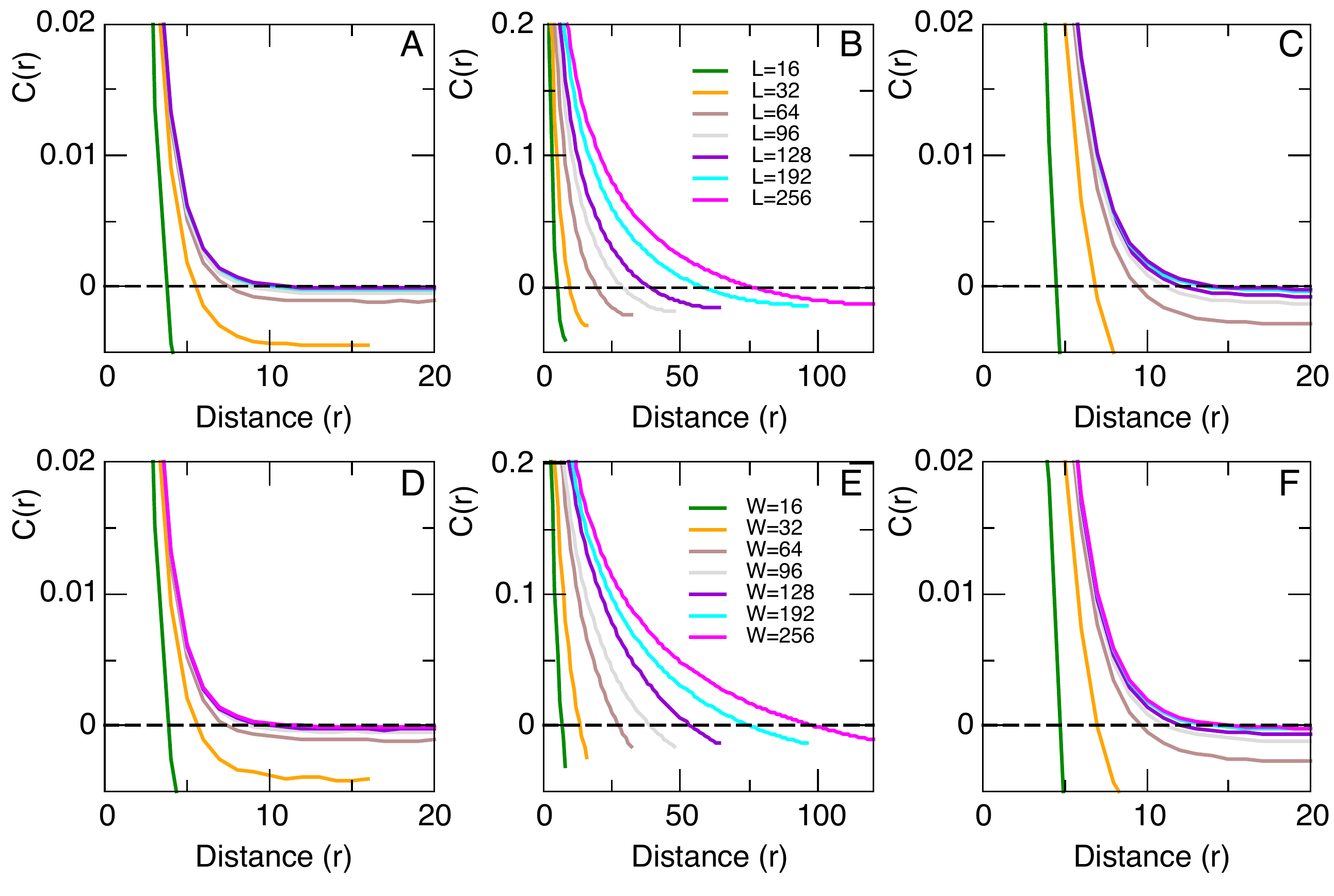} 
 \caption{\textbf{Ferromagnetic 2D Ising model. Connected correlation function:} Typical results for three temperatures  $T =2.00$ (panels A, D);  $T=2.27$ (panels B, E) and $T= 3.0$ (panels C, F) and various lattice and window sizes. Curves in the top panels computed from different system sizes $L$ and those in the bottom panels computed on a system of size $L=600$ using the window sizes $W$ indicated in the legend.}
 \label{Ising1}
\end{figure*}
%%%%%%%%%%%%%%%%%%%%%%%%%%%%%%%%%%%%%%% 
 
%%%%%%%%%%%%%%%%%%%%%%%%FIGURE 4%%%%%%%%%%
\begin{figure*}[ht!]
\includegraphics [width = .62 \linewidth] {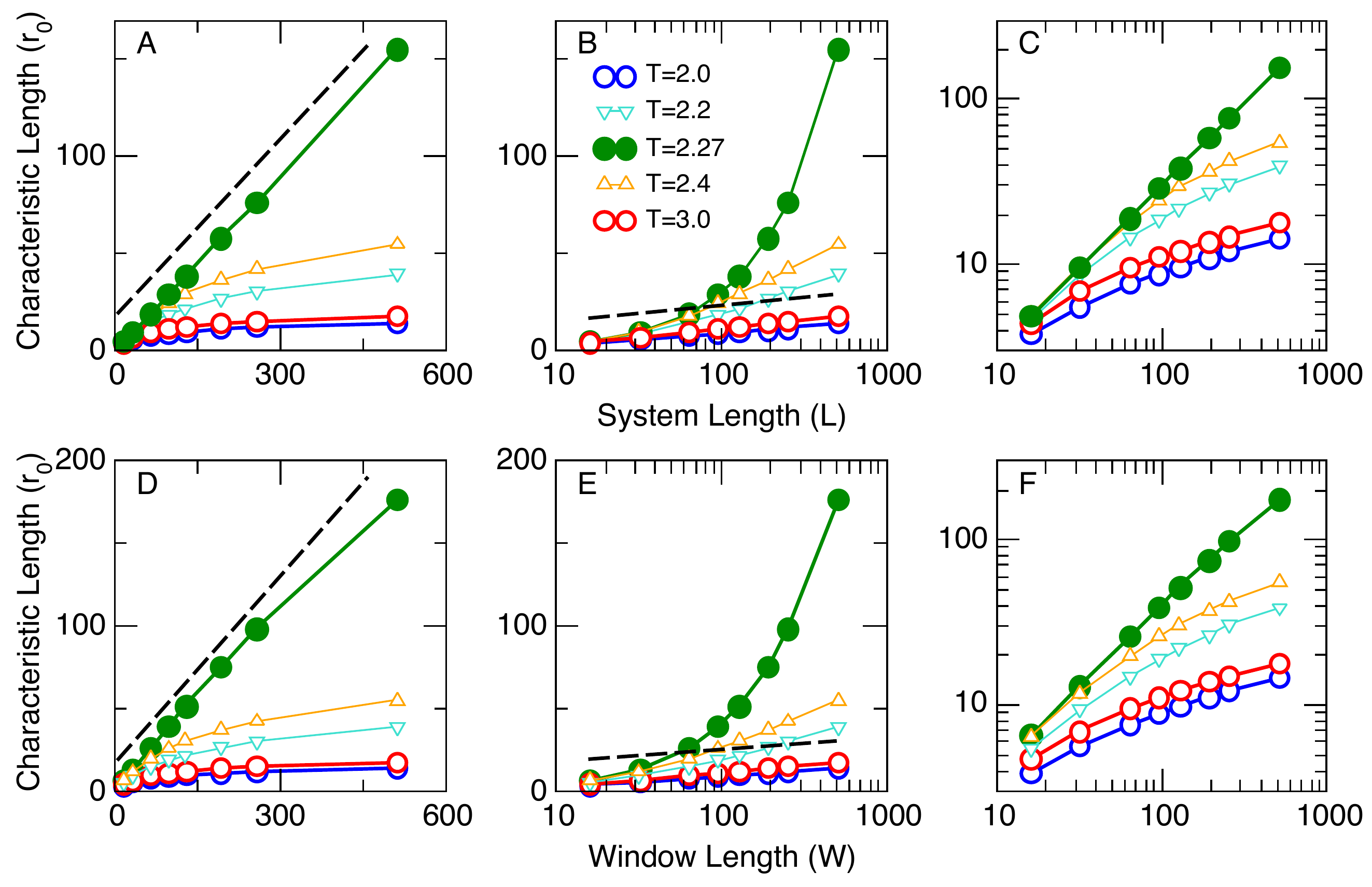} 
\caption{\textbf{Ferromagnetic 2D Ising model. Characteristic Length $r_0$.} The zero crossings of the computed CCF are plotted in linear-linear (left), log-linear (middle) and log-log (right) axis. Top three panels correspond to different system sizes $L$ and bottom three panels to different window lengths $W$. Different symbols correspond to the values of the temperature $T$ denoted in the legends. Dashed lines are visual aids to emphasize the predicted logarithmic behaviour for both sub-critical and super-critical regimes (open circles), and the linear dependence expected for the critical regime (filled circles). Open triangles are used to denote results obtained for intermediate values of temperature indicated in the legend. 
}
\label{Ising2}
\end{figure*}

The hypothesis tested here is that the behavior of $r_0$ when $W$ is varied with $L$ fixed is the same as would be obtained with $W=L$ and varying $L$, at least when $W\ll L$, namely
\begin{equation}
  \label{eq:1}
  r_0 \sim 
  \begin{cases}
    \xi \log(W/\xi), & \xi\ll W<L,\\
    W, & \xi\gg L\gg W.
  \end{cases}
\end{equation}
This behavior can be justified for physical systems in equilibrium by extending the arguments of~\cite[Sec.~2.3.3]{flocks} to the box-scaling case (see Appendix).

In the following we will study  the scaling behaviour of the characteristic length $r_0$ in two models: a 2D neuronal network as well as  in the 2D ferromagnetic Ising model.
In all simulations, $C(r)$ was measured for all integer values of $r$. Subsequently,  the smallest value of $r$ for which $C(r)$ is negative, $r_m$, was computed, and $r_0$ was found as the zero crossing of the linear fit of $C(r)$ between $r=r_m -1 $ and $r=r_m$.

\emph{{2D Neuronal network model:}} We study a neuronal network  described previously \cite{Kinouchi,tiago2014}. In short, the model is a cellular automata network on a square lattice, in which  each neuron can be in one of three states at each time step: 0 for resting, 1 for active (lasting one time step), and 2 for refractory (lasting two time steps). Each neuron connects to $K$ other neurons (here always $K=16$), in which Euclidean nearest neighbour neurons are favoured by an exponential decaying function of the distance $r$ between them (fixed here to $r_d = 5$). 
To ease the interpretation we imposed  a cutoff in the interaction probability by preventing neurons to be connected at distances $r$ greater than a given value (called here interaction length $I$, fixed here at $r=20$). 

The network overall rate of activity is set by a very small ($h = 10^{-7}  step ^{-1}$) independent Poisson perturbation to each neuron. We have verified that the results are robust over a wide range of $h$ values (e.g., $h=10^{-9}$ to $10^{-4}  step ^{-1}$).   
The model includes a control parameter  $\sigma = K \times T$,  in which $T$ is the probability that an active neuron (i.e.\ in state 1) can excite one of the $K$ neighbors that are connected to it. Therefore, as shown previously \cite{Kinouchi}, for any given value of $K$,  the model can be made critical by changing the transmission probability $T$ such that $\sigma =1$. 
We study two different scenarios: in the first we compute the CCF of the neuronal activities  collected from systems of increasing sizes, from $L=40$ up to $L=640$. In the second case, a system of (fixed) large size ($L=1000$, i.e., $L \times L$ neurons) was simulated, from which the activity of each neuron inside of windows of sizes smaller than $L$ (from $W=40$ up to $W=640$) were extracted for the correlation analysis. In all cases  we considered periodic boundary conditions.  Results presented below correspond to averages of twenty realizations lasting $2.5 \times 10^4$ steps for each parameter value.  

There are four scale lengths to consider in this model: The first is the interaction length $I$, it is the scale at which neurons interact via direct connections. The second is the system size $L$. The third is the size of the observation window ($W < L$) which determines the subset of neurons selected to compute the CCF. The last scale is the characteristic length $r_0$  which will be determined from the CCF.

Fig. \ref{F1} shows the connected correlation functions computed for various system and window  sizes, considering $s=1$ in Eq. 1, if neuron is active or refractory, and 0 otherwise, and three values of the control parameter $\sigma$. 
The results correspond to representative values of the control parameter: sub-critical $\sigma =0.64$, super-critical ($\sigma=1.6 $) and critical ($\sigma \sim 1$) as indicated in the legend. Top panels (A, B and C) correspond to computations from increasing system sizes. Bottom panels (D, E and F) correspond to CCF computed using various window length sizes  from a system of size $L=1000$.  It can be seen that the functions obtained  when changing system size $L$ or changing window size $W$ are qualitatively very similar. In particular we note that, as expected from Eq.~2, for both sub-critical  and super-critical values of $\sigma$ (panels A, D and C, F respectively) correlations do not grow much beyond the model interaction length $I=20$. At criticality, on the other hand  $r_0$ increases when either the system itself (panel B) or  windows increase in size (panel E).

The values of $r_0$ extracted from the curves in  Fig. \ref{F1} (as well as two other $\sigma$ values) are plotted in Fig. \ref{F2}. Here the same data is presented using different axis formats to visualize the different functional dependency near and away the critical point. Dashed lines in panels B and E are a guide to the eye illustrating the expected logarithmic behaviour of $r_0$ for sub-critical and super-critical regimes (open circles). The dashed lines in panels A and D  denote the linear dependence expected for $r_0$ (filled circles) in the critical regime. Finally, the same data is plotted in log-log axis in panels C and F to reveal the crossover behaviour for $W$ values close to and smaller than the interaction length ($I=20$), denoted by the deviation from the  asymptotic linear dependency for large $W$.  
Also it is worth to notice the small deviation of the linear scaling observed at criticality when the value of $W$ approaches the system size $L$ ($W=640$ in panel D of Fig~\ref{F2}). This deviation is expected from the theory, as discussed in the Appendix.  

Overall, these results show that the scaling of the characteristic length $r_0$ follows a similar functional dependence with either the  box-scaling or the system-size.

\emph{{2D ferromagnetic Ising model:}} 
The results obtained from the neuronal model were replicated in numerical simulations of the  ferromagnetic 2D Ising model.  Similar to the previous model, the simulations used two scenarios: in the first  the CCF was computed in the standard way from a model running on square lattices of increasing sizes from $L=16$ up to $L=512$. In the second setup, a relatively large $L=600$ square lattice (i.e. $600 \times 600$ spins) was simulated, and the CCF was computed from  square window  of smaller sizes from $W=16$ up to $W=512$. Results correspond to averages of five realizations each one lasting at least $5\times 10^6$ Monte Carlo steps. In all cases  we considered periodic boundary conditions.

Fig. \ref{Ising1} shows representative results for the two scenarios and three different temperatures: sub-critical ($T=2.0$, Panels A and D), critical ($T=2.27$, Panels B and E) and super-critical ($T=3.0$, Panels C and F). The top panels represent results computed for increasing system sizes and the  bottom panels for a fixed lattice size and various window sizes.
Note that, as already seen in the simulations of the neuronal model, the computation of the CCF  by changing system size $L$ or by changing window size  $W$ produces very similar results. 

The dependence of $r_0$ with system and window size is shown in Fig.~\ref{Ising2} using the same format as in Fig.~\ref{F2}  for the neuronal model. It is clear that the results obtained from varying the system or the window size show a striking similarity,  suggesting that for this system the approximation is also valid. Notice the small deviation from linearity observed at criticality for $W=512$ in panel D of Fig.~\ref{Ising2} which is similar to that exhibited by  the neuronal model for $W$ sizes near the value of $L$. 

In summary, the results obtained from a neuronal network model and from the ferromagnetic 2D Ising model  show that
the finite-size scaling of the correlation length $\xi$ can be approximated ---near the small-size limit--- by the dependence  of the characteristic length $r_0$ on window size.  The results are particularly relevant at the experimental level in neuroscience, in which techniques to map different areas of the brain cortex are now available~\cite{hollopaper}, while changing system size is not feasible.  In that  direction, the present analysis is fully consistent with the experimental observations being reported in~\cite{tiago2020}.
  
\emph{Acknowledgements:} Work supported by 1U19NS107464-01 BRAIN Initiative (USA), ZIA MH00297 of the DIRP, NIMH (USA) and CONICET (Argentina). DAM acknowledges  additional support from ANPCyT Grant No. PICT-2016-3874 (AR).

\bigskip

\def\CP{C_\text{ph}}

\textbf{APPENDIX: Box-size scaling in equilibrium.}  The relationship between $r_0$ and $W$ observed in our numerical experiments can be understood in the case of a physical system in equilibrium, where one can relate the CCF computed with space averages (as we do here) to the usual phase-average connected correlation.  We show here how to adapt the argument of \cite[Sec.~2.3.3]{flocks} to the case of $W<L$.  We start with the relationship between $C(r)$ and the correlation $\CP(r)$, defined as $C(r)$ but using the fluctuations with respect to the phase average $\langle s_i\rangle$, which for $W\gg a$ (where $a$ represents the microscopic lengths such as lattice spacing or interaction range) is \cite{flocks}
\begin{equation}
  \label{eq:2}
  C(r) = \CP(r) - \left\langle [ \overline{s} - \langle\overline{s}\rangle ]^2 \right\rangle.
\end{equation}
The variance of $\overline{s}$ computed over a volume $W^d$ can be written in terms of $\CP(r)$,
\begin{equation}
  \label{eq:3}
  \left\langle [ \overline{s} - \langle\overline{s}\rangle ]^2 \right\rangle = \frac{1}{W^d} \int_{W^d} \!\! d^dr\, \CP(r) g(r),
\end{equation}
where $g(r)$ is the radial distribution function describing the density correlations of the lattice, and arises here because the definition of the CCF includes a denominator which is essentially $r^{d-1} g(r)$.  The definition of $r_0$ is $C(r_0)=0$, so that
\begin{equation}
  \label{eq:4}
  \CP(r_0) = \frac{1}{W^d} \int_{W^d} \!\! d^dr\, \CP(r) g(r).
\end{equation}
This equation is useful because, for equilibrium physical systems near a critical point, we know the scaling form of $\CP(r)$, which we can use to obtain the relationship we seek.

We must distinguish two cases:

\emph{(i)} $\xi \ll W<L$: In this case we can write for $r\gg a$,  $\CP(r) = r^{-d+2-\eta} e^{r/\xi}$.  At this scale the system is homogeneous, and we can approximate $g(r)\approx 1$. Clearly $r_0$ will depend on $\xi$ and $W$ but not on $L$.  Due to the short range of $\CP(r)$ the integral in Eq.~\ref{eq:4} can be extended to infinity, so that
\begin{equation}
  \label{eq:5}
  r_0^{-d+2-\eta} e^{-r_0/\xi} = \frac{1}{W^d} \xi^{2-\eta} \int^\infty \!\!dx\, x^{1-\eta} e^{-x},
\end{equation}
which gives to leading order
\begin{equation}
  \label{eq:6}
  r_0 \sim \xi\log W.
\end{equation}

\emph{(ii)} $\xi\gg L \gg W$.  This is the critical case, where $\CP(r) = r^{-d+2-\eta} h(r/L)$ for $r\gg a$ \cite{Cardy}.  For $L\to\infty$, the scaling function $h(x)$ goes to a constant and the decay is a pure power law, but for finite $L$ the decay is modulated by the scaling function.  Plugging into Eq.~\ref{eq:4} and using again $g(r)\approx1$ we get
\begin{equation}
  \label{eq:7}
  r_0^{-d+2-\eta} h(r_0/L) = W^{-d} L^{2-\eta} \int^{W/L} h(u) u^{1-\eta}\,du.
\end{equation}
If $W=L$ the integral reduces to some constant, and we see that $r_0 \sim L$ is a solution, justifying the claim that the zero of $C(r)$ is proportional to $L$ when the correlation is computed over the whole sample.  If $W<L$ (i.e.\ if $C(r)$ is computed over a box smaller than the whole system), then in general $r_0$ will depend on both $L$ and $W$.  However if $W\ll L$ we are in a regime where $C(r)$ should decay almost as a pure power law, because the modulating effects of $h(x)$ will be noticeable only for $r \approx L$.  This means that we can replace $h(u)$ with a constant inside the integral, so that
\begin{equation}
  \label{eq:8}
  r_0^{-d+2-\eta} h(r_0/L) \sim W^{-d} L^{2-\eta} \int^{W/L} u^{1-\eta}\,du \sim W^{-d+2-\eta},
\end{equation}
which gives $r_0\sim W$.

\end{document}